\documentclass{iaus}
\usepackage{graphicx}
\usepackage{url}

\title[The impact of Gaia and LSST on binary stars and exo-planets] 
{The impact of Gaia and LSST on binaries and exo-planets}

\author[Laurent Eyer et al.]  
{L. Eyer$^1$,
  P. Dubath$^2$, 
  N. Mowlavi$^2$,
  P. North$^3$,
  A. Triaud$^1$,
  F. Barblan$^1$,
  C. Siopis$^4$,
  L. Guy$^2$,
  B. Tingley$^5$,
  S. Zucker$^6$,
  D.W. Evans$^7$,
  \L. Wyrzykowski$^{7,8}$,
  M. S\"uveges$^2$
 \and Z. Ivezic$^9$
 }

\affiliation{
$^1$Geneva Observatory, University of Geneva, Sauverny, Switzerland \\ 
$^2$ISDC, Geneva observatory, University of Geneva, Versoix, Switzerland\\
$^3$LASTRO, Ecole Polytechnique F\'ed\'erale de Lausanne (EPFL), Observatoire
de Sauverny, Versoix, Switzerland\\
$^4$IAA, Universit\'e Libre de Bruxelles, Bruxelles, Belgium\\
$^5$Instituto de Astrofísica de Canarias, La Laguna, Spain\\
$^6$Department of Geophysics \& Planetary Sciences, Tel Aviv University, Tel Aviv, Israel\\
$^7$Institute of Astronomy, University of Cambridge, UK\\
$^8$Warsaw University Observatory, Warsaw, Poland\\
$^9$Department of Astronomy, University of Washington, Seattle, USA \\
}

\pubyear{2011}
\volume{282}  
\pagerange{119--126}
\jname{From Interacting Binaries to Exoplanets:
Essential Modeling Tools}
\editors{A.C. Editor, B.D. Editor \& C.E. Editor, eds.}
\begin{document}

\maketitle

\begin{abstract}



Two upcoming large scale surveys, the ESA Gaia and LSST projects, will bring
a new era in astronomy. The number of binary systems that will be observed
and detected by these projects is enormous, estimations range from millions
for Gaia to several tens of millions for LSST.  We review some tools that should
be developed and also what can be gained from these missions on the subject of
binaries and exoplanets from the astrometry, photometry, radial velocity and
their alert systems.

\end{abstract}
\firstsection 
\section{Introduction}
What new knowledge will large surveys such as Gaia and LSST bring to the field of binary stars and exoplanets?
Astronomy has evolved into a science of large numbers. With Gaia and LSST, this trend is even accelerating. With these very large numbers of observed objects, it will be possible to (a) describe statistically populations of stars, binary/multiple stars and exoplanets, revealing relations between their physical properties and (b) find very rare objects among the many millions, which may shed light on specific physical processes or capture these objects in a very brief moment of their evolution.

Complementary to the technological challenges that these ambitious projects require to fulfill their stringent requirements, there are software tools which have to be developed to handle the vast amount of data, to compute, search, classify and browse through Terabytes or even Petabytes of data. 

To predict the scientific impact of a project on a specific topic can be a very perilous exercise, especially when the project observes domains of astrophysical parameters never explored before. We therefore proceed with caution, starting with a description of the projects and their characteristics, followed by a brief overview of some analysis tools for characterization and classification, and finally we present a general description of binary stars and exoplanets.

\section{The Gaia and LSST projects}
{\bf Gaia:} is a space mission of the European Space Agency, which will be located at the L2 point, 1.5 million km from Earth. It will observe all stars between mag $V\simeq$6 to 20, amounting to about 1 billion objects. The measurements consist of astrometric, photometric, spectrophotometric and spectroscopic data. The length of the mission is 5 years with a possible one year extension. For 5 years, the average number of measurements will be about 70 per object. The launch is foreseen in 2013. There will be an alert system and intermediate data releases throughout the mission. The final results will be made available by 2020-2021. 

{\bf LSST:} is a ground-based telescope that will observe about half of the sky, with a harvest amounting to 10 billion stars and 10 billion galaxies. The measurements will consist of repeated positions and photometry in the 5 Sloan bands plus the ``y'' band. The survey ranges from mag $r\simeq$16 to 24, and can be extended to mag 27 with the stacking of images.
The length of the project is planned to be 10 years during which about 1000 visits will be taken for a given region of the sky. First light is planned to be in 2018. 

For any astronomy project, the constraints on space-based missions are different from those of ground-based projects. One of the bottlenecks from space is the transmission of data back to Earth. For Gaia, once a star is detected, there is an on-board data processing to compress the information into a line spread function. This ladder information is transmitted to the Earth. Consequently Gaia does not have the full pixel images. With the ground-based LSST project, all the pixel images will be stored. Other constraints from Space are the weight and size, the Gaia mission was downsized in order to fit on a Soyuz rocket, which is less costly than an Ariane launch. The advantages of space is that we get rid of the atmosphere and there is access to full sky from one instrument.

\begin{figure}[b]
\begin{center}
\includegraphics[width=13.5cm]{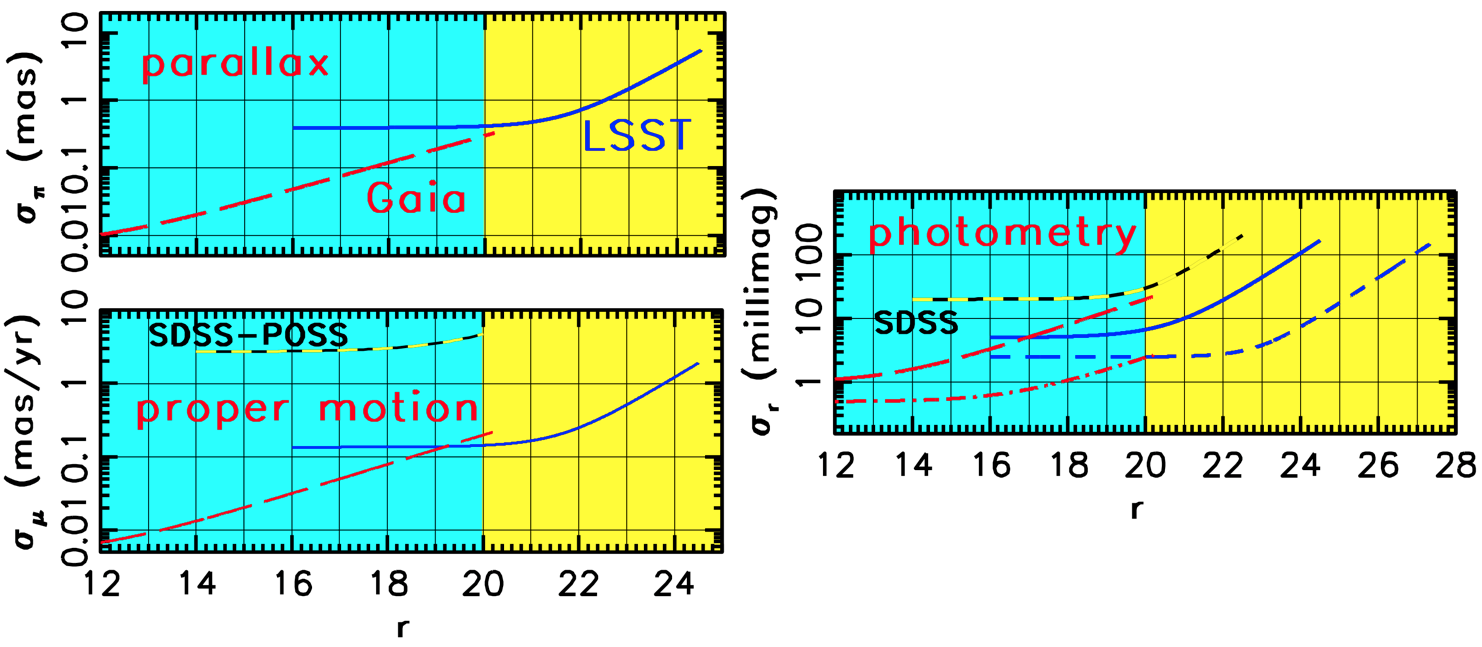}
 \caption{Precision of Gaia and LSST for the astrometry (parallax and proper motion) and photometry as function of the $r$ magnitude. In red (light grey) Gaia, in blue (dark grey) LSST. For the panel on photometry, two lines are drawn for each project, one is the precision for ``one'' measurement ($=$~a transit for Gaia, i.e. one passage on the 9 (or 8) CCDs of the Astrometric Field (AF); $=$~a ``visit'' for LSST, i.e. a pair of 15-second exposures), the others (lower) lines are the end-of-mission precision. The dashed colored lines are for SDSS and SDSS-POSS data.}
   \label{fig:precision}
\end{center}
\end{figure}

A comparison of the instruments of these two contemporary projects, Gaia from space, and LSST on Earth, are presented in Table~\ref{tab:gaialsst}. For the performance we refer to Figure~\ref{fig:precision}, which presents the astrometric (parallax, proper motion) and photometric performances. In addition to photometric measurements, Gaia has a spectrograph with a resolution 11,500 which will provide radial velocities up to magnitude $V\simeq$17 at the level of 1-10 $km s^{-1}$ depending on the spectral type of the star.

\begin{table}
  \begin{center}
  \caption{Technical characteristics of Gaia and LSST telescopes.}
  \label{tab:gaialsst}
 {\scriptsize
  \begin{tabular}{|l|c|c|}\hline 
{\bf }                            & {\bf Gaia}  & {\bf LSST}                 \\ \hline
Mirror size                       &1.45mx0.45m  & diameter:8.4m              \\
Field of view (deg$^2$)           &    0.7x0.7  &   9.6                      \\ 
Point Spread Function (arcsec$^2$)&   0.14x0.4  &   0.7x0.7   \\
Pixel counts (billion)            &          1  &   3.2                      \\ \hline
sky coverage of survey            &  whole sky  &   half of the sky          \\ 
Depth per Observation             & $V\simeq$20  &   $r\simeq$24                 \\
Bright limit                      &  $V\simeq$6  &   $r\simeq$16-17                    \\ 
Number of epochs per object       &70 in 5 years&   1000 in 10 years         \\
Number of photom. obs. per object &   70*4=280  &   1000                     \\
Number of RVS obs. per object     &         40  &    --                      \\ \hline  
  \end{tabular}
  }
 \end{center}
\end{table}

Descriptions of the science cases can be found for Gaia on the Gaia webpage at ESA   
(\url{www.rssd.esa.int/Gaia} under information sheets) and for LSST in \cite{2009arXiv0912.0201L} (or the \url{www.lsst.org}).

\section{Automated variable source characterization and supervised classification in large-scale surveys}

Source classification is an important component of any large-scale survey. The resulting large sample of variable stars is useful both to study stellar population properties and to provide candidates for further detailed investigation of individual cases. This is particularly true for binaries and exoplanets. The numbers involved in recent surveys are however so large that this task requires the development of automated and efficient machine-learning techniques.
Different supervised classification schemes are evaluated on controlled samples of well-known stars using cross-validation approaches of two recent studies (\cite[Dubath et al. 2011]{Dubathetal2011} and \cite[Richard et al. 2011]{Richardsetal2011}), each carried out in the framework of preparatory work for the Gaia and the LSST missions respectively. Both studies acknowledge the quality and convenience of the random forest method (\cite[Breiman 2001]{Breiman2001}), which is shown to perform as well as  other competing techniques. As several techniques lead to comparable results, it seems that important improvements are unlikely to come from changes in the classification method itself. The choice of the attributes used in the classification is more critical. They have to describe the source in the most appropriate manner and adding well-designed attributes, introducing a complementary piece of information if possible, is more likely to result in positive changes. The use of principal component analysis with a multicolour photometry (e.g. in \cite[S\"uveges et al. 2011]{suvegesetal2011}) is an attempt to use efficiently the variability in different photometric bands.

The quality and, above all, the relevance of the training sample is another critical factor. Any bias between the training set and the to-be-classified samples is likely to be reflected in the classification results. Additional work is currently being pursued to investigate how to best adapt a given training set to a particular survey.

\section{Binaries}
For those who work on large-scale surveys, binary stars are a problem
due to the complexity and diversity of their signals. This complexity
is also what makes them interesting. Consequently, significant effort 
is devoted to signal processing and analysis  to extract binaries from
large-scale surveys and to derive their orbital parameters as well as the
astrophysical parameters of each stellar component.

Let us go back to some basic principles: binary systems can be detected and
characterized through astrometry, photometry or spectroscopy. We review in
a few words these three ``types'' of binaries and what can be derived from the
observations:

\begin{itemize}
  \item {\bf Astrometric binaries:} A binary system is called astrometric
  when the positions on the sky of its components (or at least of the primary) change
  due to their orbital motion. When the astrometric orbits are
  determined for both components, we can get the mass ratio and the inclination,
  in short, nearly all parameters of the binary system, except for the actual size
  of the orbit and the astrophysical parameters. Furthermore, if the distance
  is known, we can then obtain the real size of the orbit and hence the total
  mass from Kepler's third law. If the distance is unknown, it is possible
  to obtain a good estimate of it by reversing the reasoning: we first
  estimate the two masses of the components using the mass-luminosity
  relation (which of course restricts this method to the main sequence), then
  we use the known orbital period and angular semi-major axis of the
  relative orbit to get the distance $d$. The precision of the latter will be
  fairly good even for a rough estimate of the total mass $M_\mathrm{tot}$,
  because $d\propto M_\mathrm{tot}^{1/3}$. This method was
  applied to Atlas (a member of the Pleiades) by \cite{Panetal2004}.
It should be noted that Gaia and LSST will provide absolute
orbits\footnote{With the wide field CCD, relative orbits are of lesser
interest}.
There are complications in some cases; for example, when the stars are not
resolved, we may only see the motion of the photocentre of the binary system.
A further complication arises in the case of a variable component in an
unresolved optical double; the photocentre will move (Variability Induced
Movers) and will thereby mimic the behaviour of a true binary system.

Within Gaia a whole group on astrometric binaries is led by D. Pourbaix
from Universit\'e Libre de Bruxelles, Belgium.
This group is dedicated to the art of solving double and multiple systems.
 \item {\bf Spectroscopic binaries:} A binary system is called spectroscopic
  when at least one component (SB1) shows a radial velocity varying
  periodically with time due to orbital motion. When lines of both components
  are seen, it is called a double-lined (SB2) system. 
  The general problem with spectroscopic
  binaries is that there is a degeneracy between orbital velocity and
  inclination $i$. With SB1 systems, we can only determine the mass function
  $M_2^3 sin^3(i)/(M_1+M_2)^2$. With SB2 systems, we get the mass ratio, but
  only a lower limit to the masses can be determined, as long as the
  inclination remains unknown.

 \item {\bf Binaries from photometry}: We can distinguish two cases, one requiring
 a time series, the other not.
  \begin{itemize}

  \item In the first case, the eclipses or tidal deformations are detected
  in the photometric time series. When only photometry is available and if
  eclipses are observed, we may get with good confidence: (a) the period;
  (b) the sum of the relative radii ($(R_1+R_2)/$(semi-major axis of the
  relative orbit)). The ratio of the radii is well constrained only if
  the eclipse is unambiguously observed as being total. With the
  relatively low number of measurements by Gaia, this will be true only for
  a fraction of all totally eclipsing binaries observed; (c) if the system
  is eccentric, $e \cos(\omega)$ can be reliably determined from the phase
  of the secondary minimum; (d) the inclination $i$ can be determined,
  provided the bottom of the minima are well observed.

  \item Binary stars can be detected through their special location in
  colour-colour or colour-magnitude diagrams.
For example, a binary sequence in a cluster can be observed in the
colour-magnitude diagram, which runs parallel to the main sequence, as
shown by e.g. \cite{Mermilliodetal1992}.
Another example is that of stars showing up in the Herzsprung gap, because
they are binaries consisting of a turn-off star and of a red giant (see e.g.
star H110 in  NGC 752, Fig.~1 of \cite[Mermilliod et al. 1998]{Mermilliodetal1998}).
Binaries can also be detected in a colour-colour diagram, as exemplified
by \cite{Smolcicetal2004}: binaries made up of a white dwarf and an M
dwarf have been detected in the SDSS database.
 \end{itemize}
\end{itemize}

Obviously, when a binary system is at the same time ``eclipsing and
spectroscopic'', it can be fully characterized (orbit, masses and radii),
see for example \cite{Barblan1998}, \cite{Northetal2010}, and when it
is ``astrometric and spectroscopic'', then the
orbit is completely determined (\cite[Zwahlen et al. 2004]{Zwahlenetal2004}).

Gaia will contribute directly to these three approaches. LSST on the 
other hand will contribute to the astrometric and photometric detections of binary
systems as well as their characterization. Obviously, Gaia and LSST will
detect objects that will prove excellent targets for follow-up studies.
The number of astrometric binaries that will be detected by Gaia and LSST
is not so well constrained.

Gaia and LSST, being extensive photometric surveys, will contribute
most significantly to the detection and characterization of new eclipsing
binaries.
Much effort is devoted to develop software to deal with this type of behaviour.
Within Gaia there is a special group, led by C. Siopis (from Universit\'e Libre
de Bruxelles, Belgium), developing 
software for the characterization of the detected eclipsing binaries. 
This software includes a number of light curve simulation and fitting 
tools which are adapted to work within mission constraints (duration, 
number of observations, software guidelines, etc.) and to make optimal 
use of mission strengths (multicolor photometry, spectroscopy, etc.).
Several attempts have been made to pin down the number of eclipsing systems
that Gaia and LSST are expected to detect. For Gaia, these estimates
disagree by a factor of nearly 10, going from half a million to 6 millions.
However, we can assert with confidence that the number of known eclipsing
systems in our Galaxy will make a prodigious jump.
Predictions for LSST have been recently made by \cite{Prasetal2011}:
the number will amount to 24,000,0000, among which 6,700,000 will be
well-characterized systems, and 1,700,000 are expected to show up as
eclisping double-lined binaries, that could be confirmed as such by follow-ups.
The largest homogeneous sample of eclipsing binaries to date comes
from the Large Magellanic Cloud OGLE-III data (\cite[Graczyk et al. 2011]{Graczyketal2011}),
from which 26,121 systems have been extracted. Thanks to Gaia and LSST,
these numbers will increase by several orders of magnitude.

There are many other subjects that can't be covered in this short review:
the case of AM CVn binary stars, potential source of gravitational waves, with extremely short periods (the Gaia per CCD photometry may be suitable for a detection of eclipses); the binary stars with a pulsating component; etc.

\section{Eruptive and cataclysmic phenomena}
Cataclysmic and eruptive phenomena are often due to interacting binaries.
Detection, classification and rapid response are crucial steps in
order that these transient objects be followed-up and studied during
unusual states. Both Gaia and LSST are developing alert systems.

Within Gaia the task to detect and deliver photometric alerts is
performed at the Institute of Astronomy of the University of
Cambridge.
From the satellite data acquisition to the ground-based first
calibration, this whole process for Gaia is not immediate. In the
worst case the received observations might be 48 hours old. In some
cases the promptest observations could be
accessed within a couple of hours. The alerts will be based on any
changes in the photometry of objects (or appearance of new ones) and
then the classification of transients will be performed exploiting all
available photometric measurements for a given object, as well as
the spectrophotometric one, which will also be available from Gaia.
This should assure relatively low rates of false positives in the
alerts stream.

The Gaia alert system group also proposes to activate and maintain a
watch list, providing data for objects known to be interesting and
important to follow-up (e.g. FU Ori-type stars), allowing 
these objects to be observed at the moment they move out of dormant
phase or undergo any unexpected change.

LSST will issue alerts based on photometric and astrometric changes, and any transient
event will be posted in less than 60 seconds (with a goal of 30
seconds) via web portals including the Virtual Observatory. LSST with
its dense time sampling and very wide photometric system, including a
u band, will be very sensitive to eruptive and cataclysmic
phenomena.

\section{Exoplanets}
Exoplanets produce astrometric signals similar to those produced by binary stars. We used the word ``similar'' because the signals or their duration usually have amplitudes much smaller than those of typical binary stars, and the detection of these signals has required very specific developments either from the instrumental or from the signal analysis point-of-view. In fact, the scientific importance of studying exoplanets has stirred much developments, which are also benefiting the binary star community.

In Table~\ref{tab:exopdetection} we summarize the way in which exoplanets can be detected and studied, and the possible contribution of Gaia and LSST.

\begin{table}
  \begin{center}
  \caption{The detection techniques used to discover exo-planets and the possible contribution from Gaia and LSST}
  \label{tab:exopdetection}
 {\scriptsize
  \begin{tabular}{|l|c|c|}\hline 
{\bf Detection Techniques} & {\bf Gaia}     & {\bf LSST}  \\ \hline
Astrometry               &  bright stars  &   --        \\
Transits                 &  bright stars  &   yes (though faint)   \\ 
Radial velocities        &     --         &   --        \\
Microlensing             &  foreseen      &   foreseen  \\ \hline
  \end{tabular}
  }
 \end{center}
\end{table}

Gaia will be helpful in many topics related to exoplanets, in particular for exoplanet detection and for the characterization of their host stars.
One of the essential points is that Gaia has a bright limit at magnitude 6, therefore allowing easy follow-up activities of potentially interesting sources with relatively small telescopes, without requiring large amounts of telescope time.
Here is a non-exhaustive list of contributions that the Gaia survey may provide:

\begin{itemize}
 \item Radius determination: One problem for planetary transit detection is the possible confusion between a secondary star and an exoplanet when the radius of the star is unknown. The luminosity of the primary star, estimated from the parallax, will enable the separation of giants from dwarfs. The spectrophotometric system may give an estimate of surface gravity and temperature and may also eliminate false detections due to the confusion between giant and dwarf parent stars. As pointed out by Triaud (these proceedings), the parallax will also lead to more accurate esimates of stellar masses and ages.
 \item Radial velocity variations: The precision of the radial velocity instrument on board Gaia, which is at the level of $km s^{-1}$, is obviously not suited for 
 the detection of exoplanets, which requires a precision at the level of 1 to 10 $m s^{-1}$. This level of $km s^{-1}$ can be used, however, for brown dwarf detection. Furthermore, the radial velocity instrument of Gaia may detect the radial velocity variations from grazing eclipsing binaries and can therefore eliminate another source of false detection.
 \item Target selection: Gaia, realizing a complete map of the sky for bright stars, will be most useful for target selection for further studies or surveys. For example, a mission such as Plato\footnote{Plato is an ESA project, which aims to detect planetary transits in bright stars} will use the information collected by Gaia.
\end{itemize}

For astrometry, the position precision required to detect planets is very demanding: currently the estimate for the bright stars along the scanning direction is at the level of 20 micro-arcseconds per Gaia-transit (average of 9 CCD positions) cf. \cite{deBruijne2009}. A small degradation of the performance will generate a significant decrease in the number of the exoplanets detected by the astrometry.
\cite{Sozzetti2011} presented the following predictions for Gaia: about 1000 planets will be detected and 400 to 500 will have their masses determined at the 10\%-20\% level.  One interesting point about the astrometric method is that it is less sensitive to the orbital inclination than the radial velocity or transit detection methods.

The case of planetary transits is somewhat controversial. Indeed, the estimations range from 0 up to 5,000-30,000 (\cite[Robichon 2002]{Robichon2002}). The matter here is to distinguish between the presence of a signal and its unambiguous detection. A knowledge of possible outlying values and technical problems is primordial in such a low signal-to-noise regime. Moreover, the bulk of the exoplanetary targets will be red dwarfs, for which little is known regarding planetary populations.

\begin{figure}[b]
\begin{center}
 \includegraphics[width=10cm]{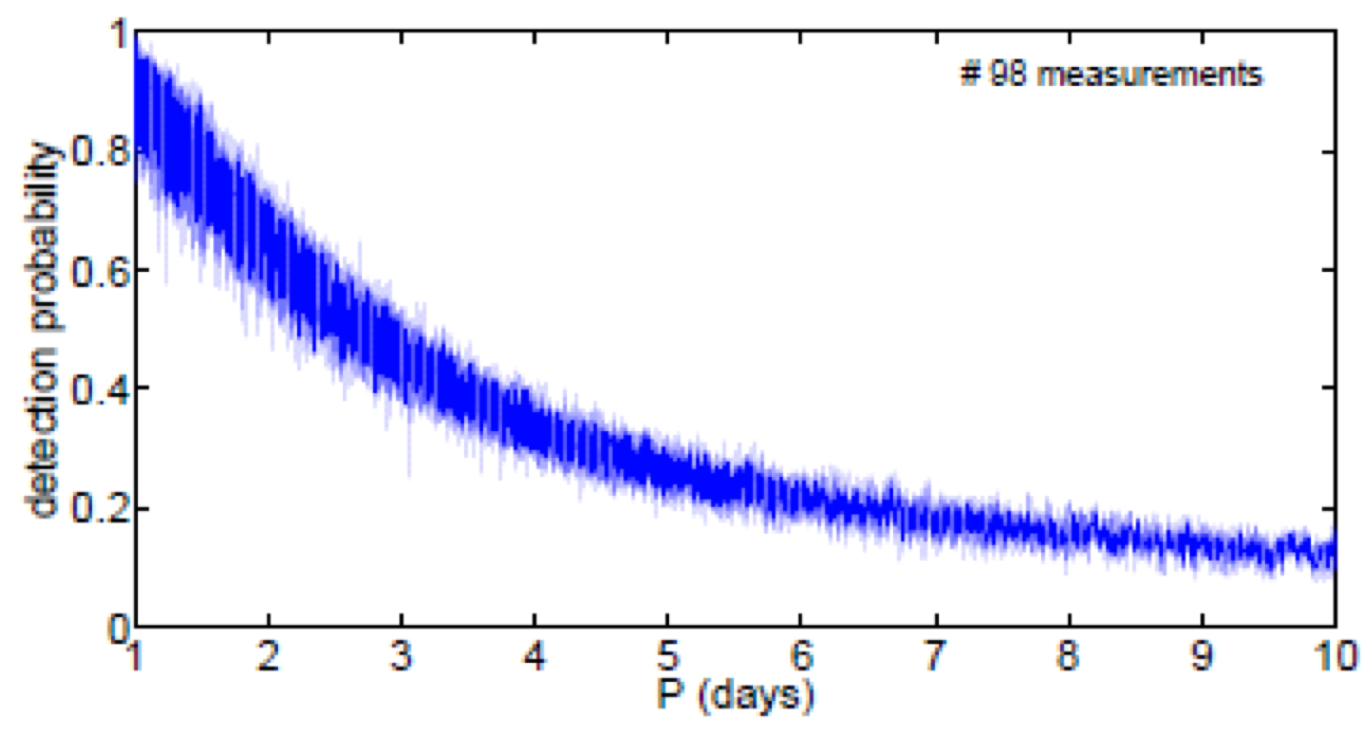} 
 \caption{TrES-1 was used as template in order to determine the probability to have 5 points in transit as a function of period (days) given the scarce Gaia sampling, here having 98 Gaia-observations over 5 years.}
   \label{fig:Prob5PinT}
\end{center}
\end{figure}

Within Gaia, \cite{Tingley2011} and \cite{DziganZucker2011} are preparing two different methods to detect such planetary signals.
Dzigan \& Zucker, using a Bayesian approach, showed that ``5 points in a transit deeper than 0.002'' means an almost certain detection. This method also allows the optimization of follow-up observations. What is then the probability that we have 5 points in transit in spite of the scarce sampling of Gaia?
Taking as an example the exoplanet TrES-1, Figure~\ref{fig:Prob5PinT} shows that this probability can be quite high. Details will be published by Dzigan \& Zucker (in preparation).

The detection of planets by microlensing is another challenge especially with Gaia because of its scarce sampling. However this activity is planned.
LSST with its denser sampling of a visit every 3 days will be more successful.

\section{Conclusions}
Gaia and LSST are two exceptional projects. If working within requirements, the scientific impact on astrophysics in general and on binaries in particular will be mind-blowing, but the significance of this impact is also difficult to predict.  This harvest does not come without effort. There are many challenges to face with such an amount of data.

This presentation may be wrong in many aspects and blind to many others. But with these caveats in mind, we can't be wrong in remarking that the quantity and quality of the data of these projects are unprecedented. 

The other lesson from this contribution is to realize that Gaia and LSST are two complementary projects. Each will cross-validate the results of the other. In addition, LSST can rely on Gaia results and extend them to fainter magnitudes.

\begin{discussion}
  \discuss{Peter Zasche}{Only a short comment. It is a pity that these new surveys have a “bright limit.”  The advantage of Hipparcos was its unlimited range in bright magnitude. Now these surveys cannot observe a 4th magnitude star. Moreover, I cannot imagine how we will observe bright stars in 20 years, when all classical photometers will be replaced with CCD camera\ldots}
  \discuss{Laurent Eyer}{When public talks are made, it is indeed somewhat embarrassing to say to the lay public: look at the night sky, and all the stars you see with your eyes, these are exactly the ones that Gaia won't measure\ldots However it concerns only a few thousands of stars, which are often too bright even for moderate size telescopes.
The magnitude ranges which are covered by Gaia and LSST are exceptional, Gaia through a gating system of the CCDs, which allows to push the bright limit to $V\simeq$6 and LSST through stacking images which allows to go very deep reaching $r\simeq$27.
Furthermore, Gaia and LSST have an extensive overlap. With these two projects we go from mag. 6 to mag. 27, which represents an exceptional dynamical range, never encountered before for such a large number of objects.}
\end{discussion}
\end{document}